\documentclass[]{interact}

\usepackage{siunitx}
\usepackage{epstopdf}\usepackage[caption=false]{subfig}
\usepackage{footmisc}

\usepackage{natbib}\bibpunct[, ]{(}{)}{;}{a}{}{,}

\theoremstyle{plain}

\theoremstyle{definition}

\theoremstyle{remark}

\usepackage{hyperref}

\usepackage{xcolor}
\usepackage{svg}

\newcommand{\rebmod}[1]{#1
}
\newcommand{\rebnew}[1]{#1
}

\begin{document}

\title{Hyperspectral Image Segmentation: A Preliminary Study on 
the Oral and Dental Spectral Image Database
(ODSI-DB)
}

\author{
    \name{Luis C. Garcia Peraza Herrera\textsuperscript{a},
\thanks{
            CONTACT Luis C. Garcia Peraza Herrera. Email: luis\_c.garcia\_peraza\_herrera@kcl.ac.uk
        }
Conor Horgan\textsuperscript{a,b},
        Sebastien Ourselin\textsuperscript{a,b},
        \\
        Michael Ebner\textsuperscript{a,b}
and
Tom Vercauteren\textsuperscript{a,b}
    }
    \affil{
        \textsuperscript{a}King's College London, UK
        \\
        \textsuperscript{b}Hypervision Surgical Ltd, UK
    }
}

\maketitle

\begin{abstract}
Visual discrimination of clinical tissue types remains challenging, with traditional RGB imaging providing limited contrast for such tasks.
Hyperspectral imaging (HSI) is a promising technology providing rich spectral information that can extend far beyond three-channel RGB imaging. 
Moreover, recently developed snapshot HSI cameras enable real-time imaging with significant potential for clinical applications. 
Despite this, the investigation into the relative performance of HSI over RGB imaging for semantic segmentation purposes has been limited, particularly in the context of medical imaging. 
Here we compare the performance of state-of-the-art deep learning image segmentation
methods when trained on hyperspectral images, RGB images, hyperspectral pixels (minus spatial context), and RGB pixels (disregarding spatial context). 
To achieve this, we employ the recently released Oral and Dental Spectral Image Database (ODSI-DB), which consists of $215$ manually segmented dental reflectance spectral images with $35$ different classes across 30 human subjects. 
The recent development of snapshot HSI cameras has made real-time clinical HSI a distinct possibility, though successful application requires a comprehensive understanding of the additional information HSI offers. 
Our work highlights the relative importance of spectral resolution, spectral range, and spatial information to both guide the development of HSI cameras and inform future clinical HSI applications.
\end{abstract} 

\section{Introduction}
\label{sec:introduction}  During an intervention, the physician or surgeon has to continuously decode the visual information into tissue types and pathological conditions. 
As a result of this process, decisions on how to continue with the intervention are taken.
Hyperspectral cameras can capture visual information far beyond the three red, green, and blue (RGB) wavelengths that the naked human eye (and common endoscopes) can perceive.
This additional data provides extra cues that facilitate the identification and characterization of relevant tissue structures that are otherwise imperceptible~\citep{Shapey2019, Ebner2021}.
To name a few examples, hyperspectral information has been used as an input to
visualize tissue structures occluded by blood \citep{Monteiro2004} or ligament tissue \citep{Zuzak2008}, display tissue oxygenation and perfusion \citep{Best2011,Chin2017}, classify images as cancerous/normal \citep{Lu2017,Fei2017,Bravo2017,Beaulieu2018}, and improve the contrast between different anatomical structures such as liver/gallbladder~\citep{Zuzak2007}, 
ureter~\citep{Nouri2016} and facial nerve \citep{Nouri2016}. The output of these systems is typically a classification score, an overlay with a segmentation, or a contrast-enhanced image. In this work, we focus on segmentation.

\rebmod{Image segmentation is a building box of many computer-assisted medical applications. }
\rebnew{In dentistry, the two most common diagnostic 
visualization strategies are visual inspection (RGB imaging)
	and X-ray imaging~\citep{Hyttinen2020}.
RGB imaging serves a multitude of purposes. To name a few, patient instruction and motivation, medico-legal reasons, treatment planning, liaison with dental laboratory, assessment of the baseline situation (when seeing a new patient), and progress monitoring \citep{Ahmad2009,Ahmad2009a}. 
	RGB imaging also provides valuable information for soft-tissue diagnostics and some surface features of hard tissue.
However, the information it captures is restricted to the capabilities of the human eye, with spectral characteristics dictated by the central wavelengths of the short, middle, and long wavelength-detecting cones in the retina ($450\,$nm, $520\,$nm, $660\,$nm).
Unlike RGB imaging, X-ray provides anatomical and pathological information on hard-tissue structures such as the teeth and the alveolar bone. This additional information comes at the expense of exposing the patient to ionizing radiation and potential risks derived from the use of contrast agents.
}

\rebnew{In contrast to RGB imaging, hyperspectral cameras allow us to capture additional information beyond the usual three RGB bands. This new set of images corresponding to narrow and contiguous wavelength bands forms the reflection spectrum of the sample (in our case, the sample is the inside of the mouth). 
Although the applications and possible benefits of hyperspectral imaging (HSI) are an active field of research \citep{Shapey2019,Manifold2021,Ebner2021,Seidlitz2022}, we foresee that patients could potentially benefit from this technology in two different ways.
First, the reflectance spectrum could be used to extract tissue properties and produce a range of pseudo-color images
that enhance the visualization capabilities of clinicians \citep{Falt2018}. For example, displaying or highlighting imaging biomarkers or clinical conditions that are barely visible or not perceivable in RGB \citep{Best2013a,Zherebtsov2019,Hyttinen2018,Hyttinen2019}.
Second, hyperspectral images could provide computer-assisted diagnosis methods with additional information that can help improve the accuracy of detecting and diagnosing lesions \citep{Boiko2019}.
The work presented in this manuscript is aimed at assessing the latter possible benefit.
}

\rebnew{The research hypothesis we are working with is that there may be perceivable differences in the reflectance spectrum of diseased tissue compared to that of healthy anatomy. 
For example, the average reflectance spectrum of all the pixels of a healthy tooth might be different to that of one affected by a certain condition (e.g. early-stage cavities).   
A preliminary step to the development of such quantitative dental and oral biomarkers is to segment the different anatomical structures accurately. In this scenario, a question that quickly arises is whether we can obtain an improved segmentation accuracy with state-of-the-art deep neural network architectures designed for 2D RGB image analysis by simply replacing the RGB input with a hyperspectral cube. Similarly, from a deep neural network design perspective, it is interesting to see how accuracy changes when we just use spectral information (i.e. when we classify pixels individually) or when we also use spatial information (i.e. when we segment images).
}
\rebnew{That is, we aim to discover how the segmentation accuracy changes when reducing the information available from $N$ hyperspectral bands to the usual three RGB bands. This is indeed one way to assess a \textit{lower bound}
\footnote{We refer to \textit{lower bound} because it is possible that by tailoring the network model (instead of using a common 2D U-Net) to the nature of hyperspectral data we would render further accuracy improvements.
	}
of the added value of hyperspectral information above and beyond RGB.
}

\subsection{Contributions}
In this work, we provide baseline results for the segmentation of the Oral and Dental Spectral Image Database (ODSI-DB) \citep{Hyttinen2020}. 
We evaluate how the segmentation performance changes when using different spatial and spectral resolutions as data inputs, guiding future developments in the field of dental reflectance and hyperspectral image segmentation. 
We provide the training code along with the models validated in this work 
\footnote{\rebmod{\url{https://github.com/luiscarlosgph/segodsidb}}.
    \label{footnote:link_to_source_code}
}.
Additionally, we propose an improved approach to reconstructing RGB images from hyperspectral raw data than that employed in ODSI-DB. This is useful to compensate for missing spectra, as it occurs in the images captured with Nuance EX camera used in ODSI-DB.
Our code to perform such conversion is also made available \footref{footnote:link_to_source_code}.

\section{Related work}
\label{sec:related_work}
Recently, a literature review on deep learning techniques applied to hyperspectral medical images has been published \citep{Khan2021}. In the following paragraph we summarise some of the most recent work involving hyperspectral images in the context of surgery, and how the performance varies across different computer-assisted applications when using hyperspectral bands as opposed to traditional RGB imaging.

In \cite{Garifullin2018}, \rebnew{the} authors segmented the retinal vasculature, optic disc and macula using $30$ spectral bands ($380$-$780$\,nm). Their results showed an improvement of $2$ \rebmod{percentage points (pp) for vessels and optic disc and $6\,$pp for the macula} when comparing deep learning (DL) models trained on hyperspectral versus RGB images. 
In \cite{Ma2019,Ma2021}, authors employ hyperspectral images for tumor classification and margin assessment. In this work, the model proposed by the authors for tissue classification on surgical specimens achieved a pixel-wise average AUC of 0.88 and 0.84 for hyperspectral and RGB, respectively. 
In \cite{Wang2021a}, the authors reported a difference of $2\,$pp in Dice coefficient for the segmentation of melanoma in histopathology samples of the skin when comparing the performance of a 2D U-net on RGB and hyperspectral images.
Similarly, \cite{Trajanovski2021} showed that the Dice coefficient for the segmentation of tongue tumors increases from $0.80$ to $0.89$ when using hyperspectral information.
Despite this recent body of work, to the best of our knowledge, there is no current benchmark on ODSI-DB, which as opposed to the previous literature, mostly targeting binary classification, has a considerably higher number of classes ($35$) and a substantial number of patient samples ($>200$).

\section{Materials and methods}
\label{sec:materials_and_methods}

\subsection{Dataset details}
\label{sec:dataset_details}
The ODSI-DB dataset \citep{Hyttinen2020} contains $316$ images ($215$ are annotated, $101$ are not) of $30$ human subjects. The $215$ annotated images are partially labelled, and the number of annotated pixels per image varies from image to image. The annotated pixels can belong to $35$ possible classes. The number of annotated pixels per class is shown in Tab.~\ref{tab:class_to_numpix} in the appendix.
ODSI-DB contains images captured with two different cameras, $59$ annotated images were taken with a Nuance EX (CRI, PerkinElmer, Inc., Waltham, MA, USA), and $156$ were obtained with a Specim IQ (Specim, Spectral Imaging Ltd., Oulu, Finland). The pictures taken by the Nuance EX contain $51$ spectral bands ($450$–$950$ nm with $10$nm bands), and those captured by the Specim IQ have $204$ bands ($400$–$1000$nm with approximately $3$nm steps). The reflectance values for the images are in the normalized range $[0, 1]$.

\subsection{Reconstruction of RGB images from hyperspectral data}
\label{sec:rgb_reconstruction}
\begin{figure}[htb!]
    \centering
	\includegraphics[width=.90\textwidth]{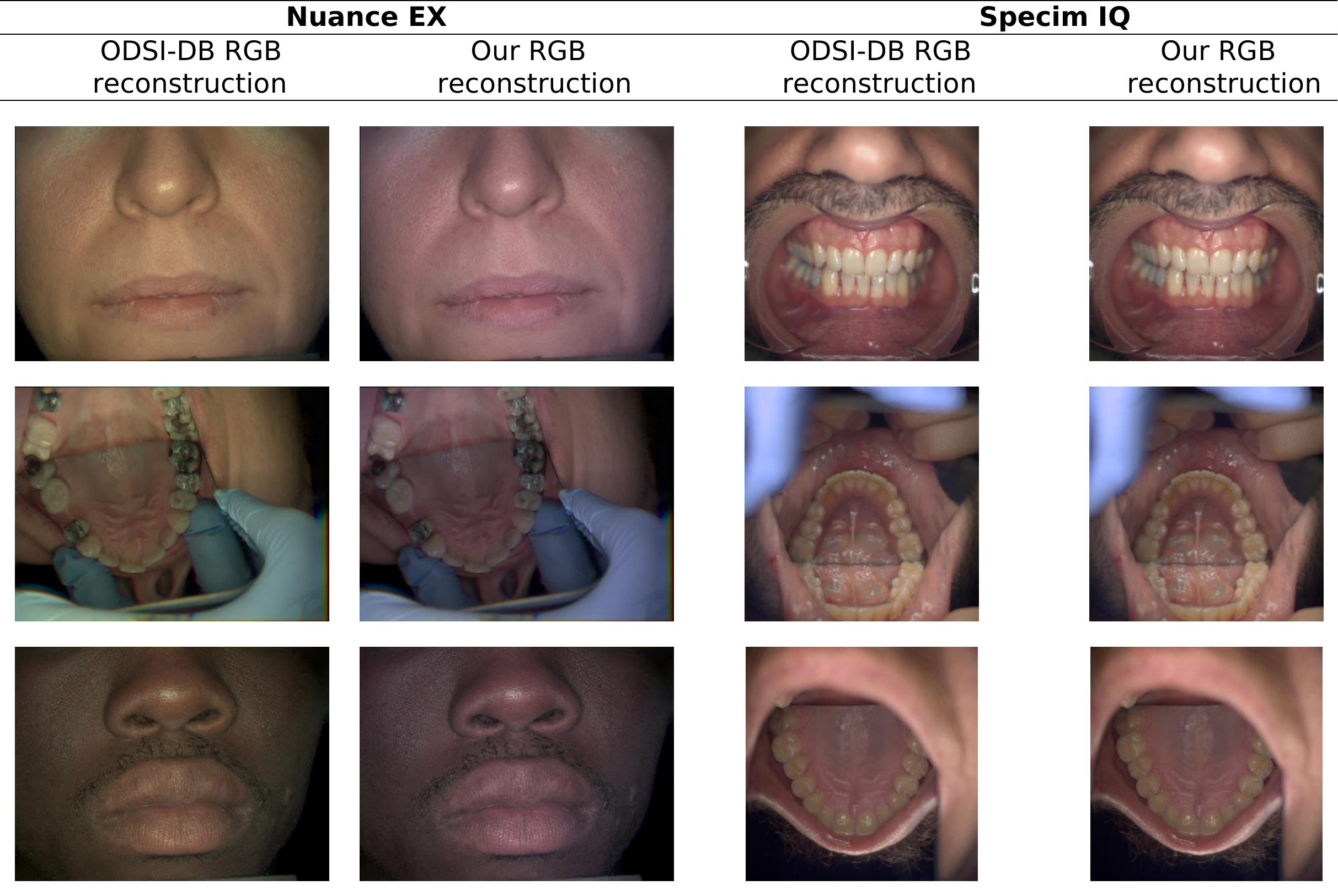}
    \caption{Exemplary reconstructed RGB images provided in the ODSI-DB dataset compared to our RGB reconstructions. As can be observed, the RGB images reconstructed from the hyperspectral images captured with the Nuance EX are affected by a yellow artifact. This artifact is not present in those reconstructed from the Specim IQ. This occurs because the Nuance EX does not capture the $400$-\SI{450}{nm} range, which carries information relevant to reconstruct the blue channel (and in a lesser degree the red) channel.
    }
	\label{fig:yellow_effect}
\end{figure}

Although ODSI-DB contains RGB reconstructions of the hyperspectral images, we have observed that the RGB reconstruction method used to generate the RGB images provided does not compensate for the lack of hyperspectral information in the $400$-\SI{450}{nm} range for the Nuance EX images. The lack of this information results in a yellow artifact in the reconstructed RGB images from the Nuance EX (see Fig. \ref{fig:yellow_effect}). We thus provide an alternative RGB reconstruction.

\rebmod{Our RGB reconstruction follows the method proposed by \citet{Magnusson2020}, where the hyperspectral images are first converted to CIE XYZ and then to sRGB.
}

\begin{figure}[htb!]
	\centering
    \includegraphics[width=0.32\textwidth]{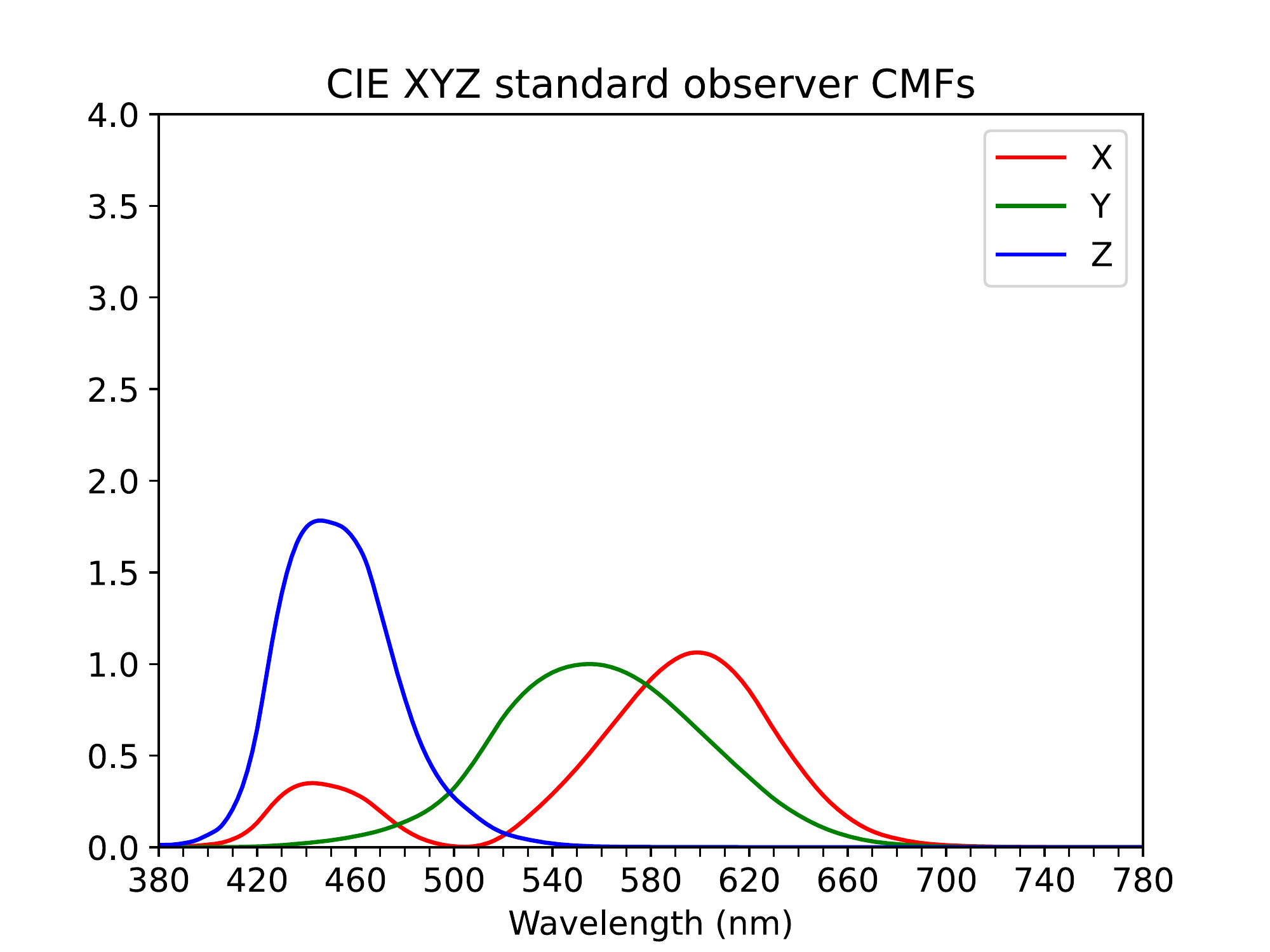}
	\includegraphics[width=0.32\textwidth]{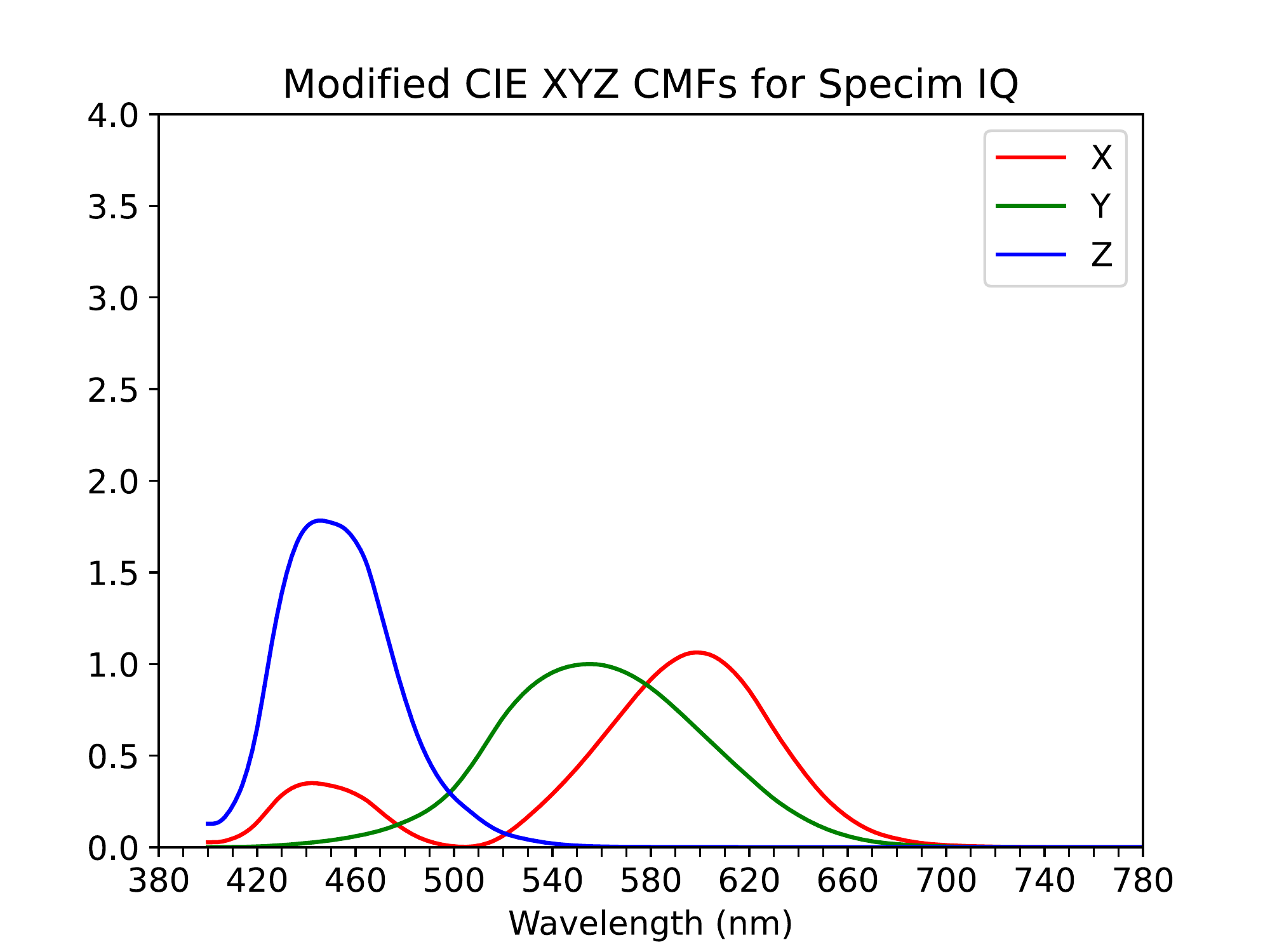}
	\includegraphics[width=0.32\textwidth]{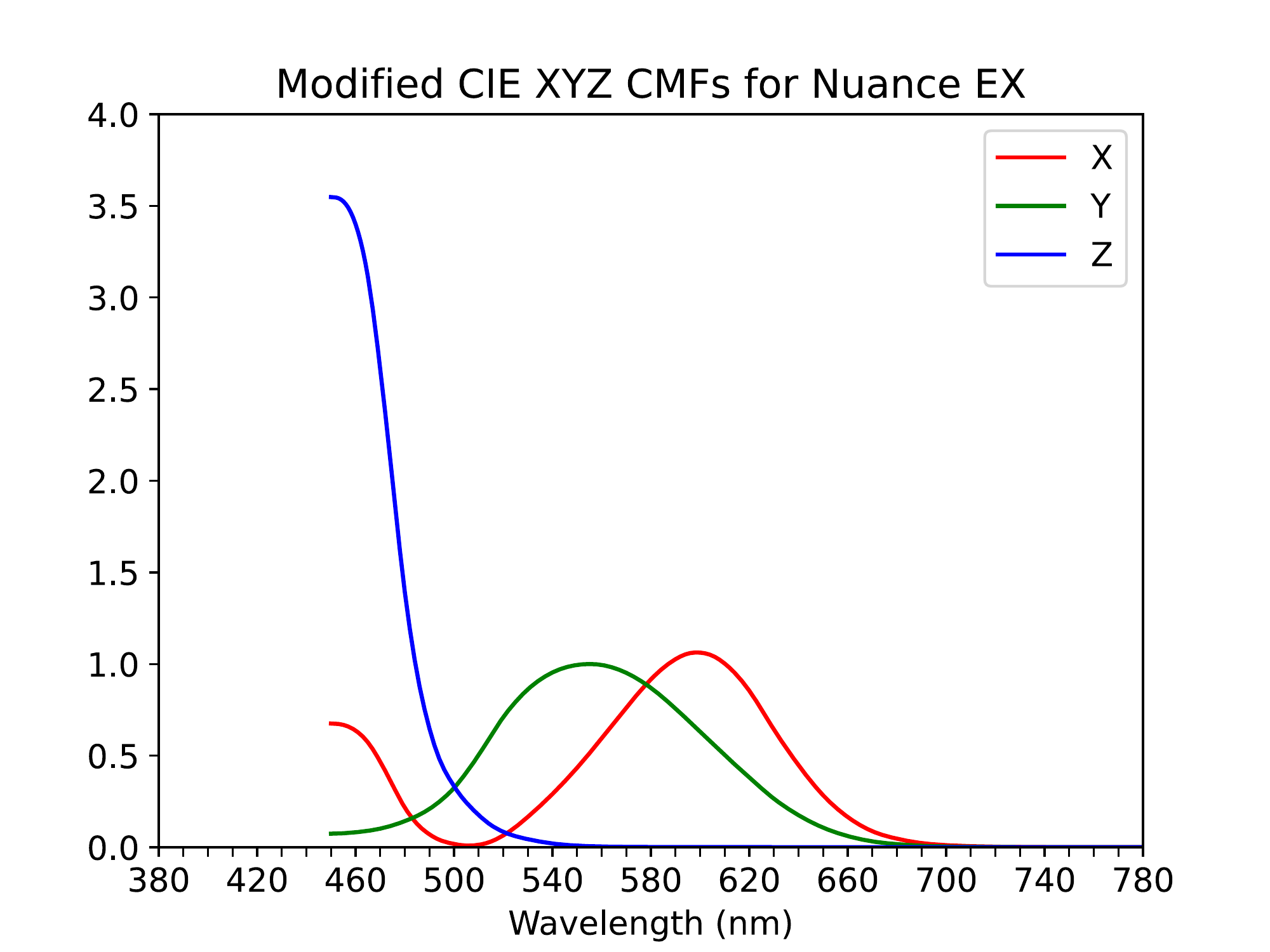}
	\caption{\rebnew{Original and modified color matching functions.
		}
	}
	\label{fig:cie_1931_xyz_cmf}
\end{figure}

\rebnew{The conversion from CIE XYZ to linear sRGB is a linear transformation where the X, Y, and Z channels contribute largely to the red, green, and blue channels of the linear sRGB image, respectively. 
When converting hyperspectral images to CIE XYZ, the contribution (i.e. the weight) of each hyperspectral band to the CIE XYZ image is defined by a color matching function (CMF). We used the standard CIE 1931, shown in Fig. \ref{fig:cie_1931_xyz_cmf}.
However, as shown in Fig. \ref{fig:cie_1931_xyz_cmf}, the hyperspectral bands in the range $400-450\,$nm have a considerable weight to reconstruct the Z channel (blue), and a minor weight to reconstruct the X (red) channel. 
\\
		\\
As the Nuance EX hyperspectral images do not have any information in this range, we miss a substantial amount of the information needed to reconstruct the Z (blue) channel correctly, which is why the images look yellow (see the blue medical-grade glove in the Fig. $1$).
On the other hand, the images captured with the Specim IQ camera have information in the $400-450\,$nm range (the camera range is $400-1000\,$nm). Therefore, the RGB reconstructions look realistic and do not display the yellow tint seen in those from the Nuance EX.
\\
	    \\
The purpose of the modified RGB reconstruction proposed in this section is to compensate for the missing information ($380-450\,$nm for the Nuance EX and $380-400\,$nm for the Specim IQ).
With this modification, we aim to make the RGB images produced from both cameras look alike prior to them being processed by the convolutional network. 
To do so and account for the missing wavelengths,  
		we modify the CIE original CMF shown in Fig. \ref{fig:cie_1931_xyz_cmf}.
The modification consists of taking the CMF function in the missing range (e.g. $380-450\,$nm for the Nuance EX), flipping it over the vertical axis at the start of the captured wavelengths ($450\,$nm for the Nuance EX), and summing it with the original CMF. More formally, the modified color matching functions are defined as
}
\rebnew{\begin{equation}
\begin{split}
	\bar{x}_{n}(\lambda) &= \bar{x}(\lambda) + \bar{x}_{c}(\lambda) \\
	\bar{y}_{n}(\lambda) &= \bar{y}(\lambda) + \bar{y}_{c}(\lambda) \\
	\bar{z}_{n}(\lambda) &= \bar{z}(\lambda) + \bar{z}_{c}(\lambda) \\
\end{split}
\end{equation}
where
$\bar{x}$, $\bar{y}$, $\bar{z}$ are the original CIE 1931 2-deg color matching functions (CMFs) \citep{Smith1931} shown in Fig. \ref{fig:cie_1931_xyz_cmf} (left),
$\bar{x}_c$, $\bar{y}_c$, and $\bar{z}_c$ are the additive corrections to compensate for the missing information, and 
$\bar{x}_n$, $\bar{y}_n$, and $\bar{z}_n$ are the corrected CMFs (shown in Fig. \ref{fig:cie_1931_xyz_cmf} center and right). 
As different cameras are missing different wavelength ranges, the additive corrections must be different. We define the CMF correction for the Nuance EX as
\begin{equation}
\begin{split}
	\bar{x}_{c}(\lambda) &= 
	\begin{cases} 
		\bar{x}(2 \times 450 - \lambda) & 450 \leq \lambda \leq 450 + (450 - 380) \\
		0 & otherwise
	\end{cases}
	\\
	\bar{y}_{c}(\lambda) &= 
	\begin{cases} 
		\bar{y}(2 \times 450 - \lambda) & 450 \leq \lambda \leq 450 + (450 - 380) \\
		0 & otherwise
	\end{cases}
	\\
	\bar{z}_{c}(\lambda) &= 
	\begin{cases} 
		\bar{z}(2 \times 450 - \lambda) & 450 \leq \lambda \leq 450 + (450 - 380) \\
		0 & otherwise
	\end{cases}
\end{split}
\end{equation}
and the CMF correction for the Specim IQ as
\begin{equation}
\begin{split}
	\bar{x}_{c}(\lambda) &= 
	\begin{cases} 
	\bar{x}(2 \times 400 - \lambda) & 400 \leq \lambda \leq 400 + (400 - 380) \\
	0 & otherwise
	\end{cases}
	\\
	\bar{y}_{c}(\lambda) &= 
	\begin{cases} 
	\bar{y}(2 \times 400 - \lambda) & 400 \leq \lambda \leq 400 + (400 - 380) \\
	0 & otherwise
	\end{cases}
	\\
	\bar{z}_{c}(\lambda) &= 
	\begin{cases} 
	\bar{z}(2 \times 400 - \lambda) & 400 \leq \lambda \leq 400 + (400 - 380) \\
	0 & otherwise
	\end{cases}
\end{split}
\end{equation}
The original CMF, along with those modified for the Specim IQ and Nuance EX images are shown in Fig. \ref{fig:cie_1931_xyz_cmf}. 
Due to the nature of the proposed CMF modifications, the modified RGB reconstruction can be seen as a color normalization that transforms the input data to a common RGB space, easing the learning of the segmentation from a dataset containing a mix of Nuance EX and Specim IQ images. 
}

\rebmod{Once we have the modified CMFs, the conversion from hyperspectral to RGB is as follows
\begin{equation}
\begin{split}
    X &= \frac{1}{N} \int_{\lambda} \bar{x}_n(\lambda) f(\lambda) g(\lambda) \,d\lambda \\
    Y &= \frac{1}{N} \int_{\lambda} \bar{y}_n(\lambda) f(\lambda) g(\lambda) \,d\lambda \\
    Z &= \frac{1}{N} \int_{\lambda} \bar{z}_n(\lambda) f(\lambda) g(\lambda) \,d\lambda \\
    N &= \int_{\lambda} \bar{y}(\lambda) g(\lambda) \,d\lambda \\
\begin{bmatrix}
    R \\
    G \\
    B 
    \end{bmatrix}
&=
\begin{bmatrix}
    + 3.2406255  &  - 1.5372080  &  - 0.4986286  \\
    - 0.9689307  &  + 1.8757561  &  + 0.0415175  \\
    + 0.0557101  &  - 0.2040211  &  + 1.0569959
    \end{bmatrix}
\begin{bmatrix}
    X \\
    Y \\
    Z
    \end{bmatrix}
\end{split}
\end{equation}
where 
$f$ is the spectral density of the sample (i.e. the continuous version of the hyperspectral image), and 
$g$ is the spectral density of the illuminant, D65 in our case.
As we have these functions ($\bar{x}_n$, $\bar{y}_n$, $\bar{z}_n$, $f$, $g$) typically sampled at different wavelengths, we interpolate all of them with a PCHIP 1-D monotonic cubic interpolator. We use the composite trapezoidal rule to evaluate the integral (with the image wavelengths as sample points).
} 

After the RGB conversion, following the proposal by \cite{Magnusson2020} to avoid images looking overly dark, we apply the \rebmod{following} gamma correction to all the RGB pixels 
\begin{equation}
    \gamma(x) = 
    \begin{cases} 
      12.92x & x \leq 0.0031308 \\
      1.055 x^{0.416} - 0.055 & otherwise
   \end{cases}
\end{equation}
where $x$ is either the red, green, or blue intensity of each pixel (correction is applied to all the RGB channels).

\subsection{Types of input and segmentation model}
\begin{figure}[htb!]
    \centering
	\includegraphics[width=.90\textwidth]{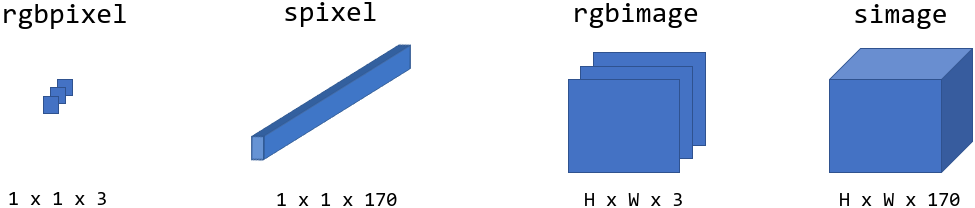}
    \caption{The four types of input compared are 
    single RGB pixels (\texttt{rgbpixel}), single hyperspectral pixels (\texttt{spixel}), RGB images (\texttt{rgbimage}), and hyperspectral images (\texttt{simage}). 
    The format used to define the dimensions of the input is $H, W, C$, where $H$, $W$, and $C$ represent height, width, and channels, respectively.
    }\label{fig:types_of_input}
\end{figure}

In this study, we evaluate model performance for pixel classification on the ODSI-DB dataset when different forms of data input are employed (see Fig. \ref{fig:types_of_input}). 

We refer to \texttt{rgbimage} when we reconstruct the RGB image from the whole spectral range using a colour matching function as explained in Sec. \ref{sec:rgb_reconstruction}.
As the hyperspectral images in ODSI-DB have a different number of bands ($450$-$950$\,nm with $10$\,nm steps for the Nuance EX, and $400$-$1000$\,nm with $3$\,nm steps for the Specim IQ), we linearly interpolate the images from both cameras to a fixed set of $170$ evenly-spaced bands in the $450$-$950$ range.
As in most of the recent body of work in biomedical segmentation, backed up by the state-of-the-art results in most biomedical segmentation challenges, we chose the endoder-decoder 2D U-Net \cite{Ronneberger2015} as our go-to model to build the segmentation baseline. For the \rebmod{pixel-wise} experiments (\texttt{rgbpixel} and \texttt{spixel}) a network with an equivalent number of $1\times1$ filters and skip layers was used. The network hyperparameters for each input type were tuned on a random $10$\% of the images contained in the training set.

\section{Results and discussion}
\label{sec:results_and_discussion}

\subsection{Hyperspectral vs RGB as feature vectors}
\label{sec:tsne}
\begin{figure}[htb!]
    \centering
	\includegraphics[width=.90\textwidth]{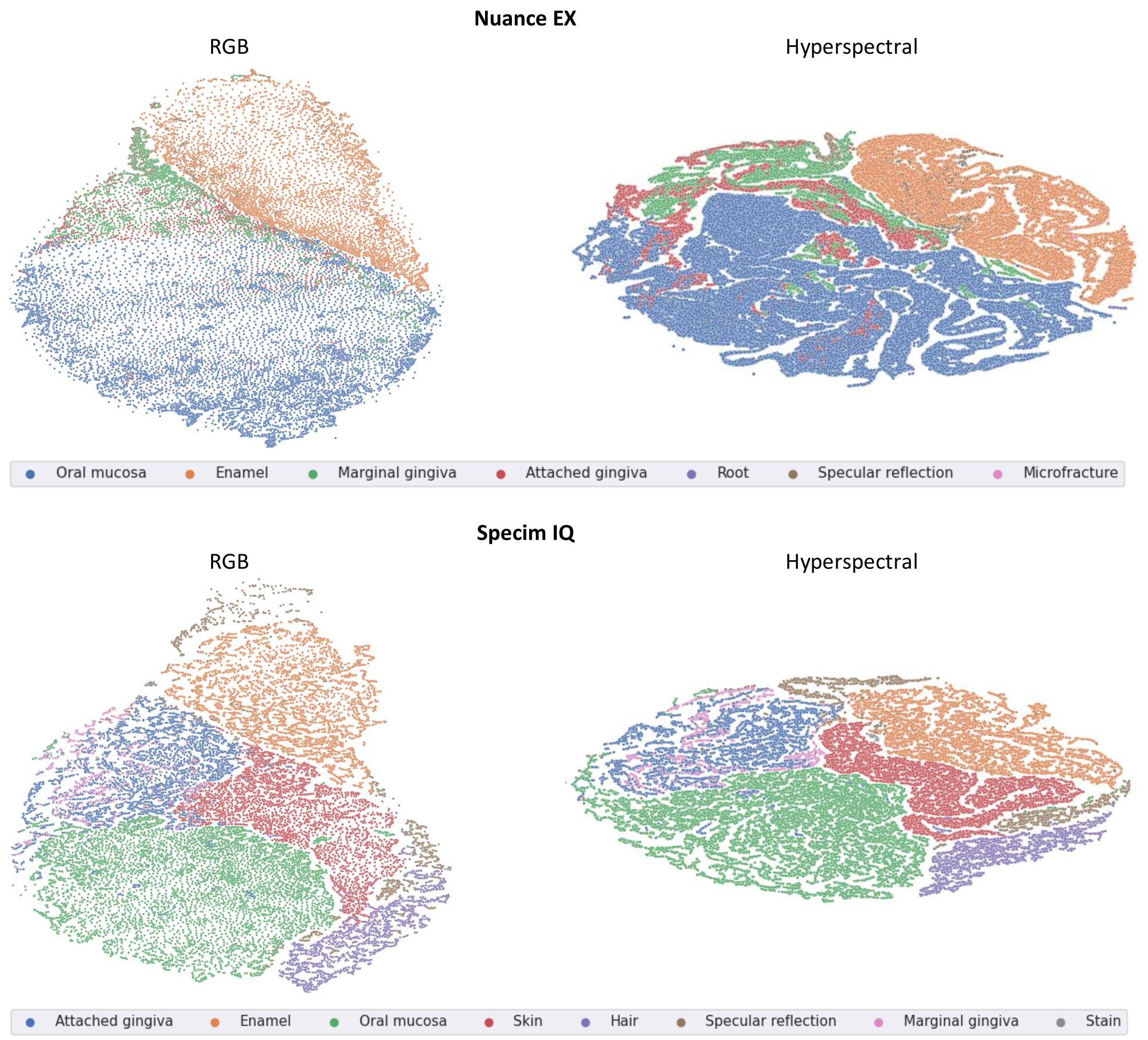}
    \caption{t-SNE of $10$ million randomly selected pixels (evenly distributed across images) from the ODSI-DB dataset. The dataset contains images captured with two different cameras, Nuance EX (51 bands) and Specim IQ (204 bands), hence the separated plots. 
    }
	\label{fig:tsne_hyperspectral_vs_rgb}
\end{figure}

Hyperspectral images have a higher spectral resolution \rebnew{than RGB images}. As we are interested \rebmod{in seeing} whether this increased resolution translates into a higher degree of discrimination among feature vectors, we run t-SNE \citep{VanderMaaten2008} on $10$ million randomly selected pixels, with an equal number of pixels picked from each image. For visualisation purposes, we reduce the dimensionality from hyperspectral and RGB to 2D. 
The hyperparameters used for the experiment were $10$, $200$, and $1000$ for perplexity, learning rate, and number of iterations, respectively. The initialization was performed with PCA. 
The visualization of the 2D features is shown in Fig. \ref{fig:tsne_hyperspectral_vs_rgb}. 
As can be observed in this figure, in the Nuance EX visualization, the hyperspectral plot shows a boundary between oral mucosa and gingiva (attached and marginal) which is blurred in RGB, and also a clearer separation between attached and marginal gingiva themselves. 
In the Specim IQ visualization, the hyperspectral plot displays a sharper edge between \textit{skin} and gingiva (attached and marginal), and also between hair and specular reflections. 

\begin{table*}[!htb]
    \centering
    \caption{Class-based pixel classification results for the \texttt{rgbpixel} mode. To generate this table all the pixels contained in the images of the testing set are considered as a single set. \rebnew{All the results are provided in percentage.}}
    \vspace{0.2cm}
    \begin{tabular}{lcccc}
        \hline
        \multicolumn{1}{c}{\bfseries Class} 
        & \multicolumn{1}{c}{\bfseries Sensitivity}
        & \multicolumn{1}{c}{\bfseries Specificity}
        & \multicolumn{1}{c}{\bfseries Accuracy}
        & \multicolumn{1}{c}{\bfseries Balanced accuracy}
        \\
        \hline
        Attached gingiva  &  54.89  &  68.08  &  67.74  &  61.49  \\
        Enamel  &  56.52  &  69.98  &  68.64  &  63.25  \\
        Hair  &  100.00  &  68.48  &  68.94  &  84.24  \\
        Hard palate  &  44.01  &  68.63  &  66.40  &  56.32  \\
        Lip  &  47.80  &  68.15  &  66.72  &  57.98  \\
        Oral mucosa  &  42.95  &  68.37  &  66.16  &  55.66  \\
        Skin  &  38.86  &  69.33  &  62.48  &  54.10  \\
        Soft palate  &  0.00  &  67.37  &  67.12  &  33.68  \\
Tongue  &  2.14  &  62.77  &  54.61  &  32.46  \\
        \hline
\multicolumn{1}{c}{\bfseries Average}   &  43.02  &  67.91  &  65.42  &  55.46  \\
        \end{tabular}
    \vspace{0.2cm}
    \label{tab:rgbpixel_results}
\end{table*}

\begin{table*}[!htb]
    \centering
    \caption{Class-based pixel classification results for the \texttt{spixel} mode. To generate this table all the pixels contained in the images of the testing set are considered as a single set. \rebnew{All the results are provided in percentage.}}
    \vspace{0.2cm}
    \begin{tabular}{lcccc}
        \hline
        \multicolumn{1}{c}{\bfseries Class} 
        & \multicolumn{1}{c}{\bfseries Sensitivity}
        & \multicolumn{1}{c}{\bfseries Specificity}
        & \multicolumn{1}{c}{\bfseries Accuracy}
        & \multicolumn{1}{c}{\bfseries Balanced accuracy}
        \\
        \hline
        Attached gingiva  &  54.73  &  68.09  &  67.74  &  61.41  \\
        Enamel  &  67.49  &  68.00  &  67.95  &  67.74  \\
        Hair  &  100.00  &  68.48  &  68.94  &  84.24  \\
        Hard palate  &  44.01  &  68.64  &  66.40  &  56.32  \\
        Lip  &  60.00  &  66.77  &  66.30  &  63.39  \\
        Oral mucosa  &  42.94  &  68.48  &  66.26  &  55.71  \\
        Skin  &  39.33  &  69.07  &  62.38  &  54.20  \\
        Soft palate  &  0.00  &  67.37  &  67.12  &  33.68  \\
Tongue  &  2.14  &  62.77  &  54.61  &  32.46  \\
        \hline
\multicolumn{1}{c}{\bfseries Average}   &  45.63  &  67.52  &  65.30  &  56.57  \\
        \end{tabular}
    \vspace{0.2cm}
    \label{tab:spixel_results}
\end{table*}

\begin{table*}[tb]
    \centering
    \caption{Class-based pixel classification results for the \texttt{rgbimage} mode. To generate this table all the pixels contained in the images of the testing set are considered as a single set. \rebnew{All the results are provided in percentage.}}
    \vspace{0.2cm}
    \begin{tabular}{lcccc}
        \hline
        \multicolumn{1}{c}{\bfseries Class} 
        & \multicolumn{1}{c}{\bfseries Sensitivity}
        & \multicolumn{1}{c}{\bfseries Specificity}
        & \multicolumn{1}{c}{\bfseries Accuracy}
        & \multicolumn{1}{c}{\bfseries Balanced accuracy}
        \\
        \hline
        Attached gingiva  &  26.23  &  99.94  &  98.00  &  63.09  \\
        Enamel  &  49.40  &  99.04  &  94.11  &  74.22  \\
        Hair  &  84.77  &  98.94  &  98.73  &  91.85  \\
        Hard palate  &  0.53  &  99.64  &  90.64  &  50.08  \\
        Lip  &  33.94  &  98.91  &  94.34  &  66.43  \\
        Oral mucosa  &  78.03  &  92.09  &  90.87  &  85.06  \\
        Skin  &  78.19  &  97.22  &  92.94  &  87.71  \\
        Soft palate  &  49.13  &  99.42  &  99.23  &  74.27  \\
        Tongue  &  56.48  &  97.20  &  91.73  &  76.84  \\
        \hline
        \multicolumn{1}{c}{\bfseries Average}   &  50.74  &  98.04  &  94.51  &  74.39  \\
        \end{tabular}
    \vspace{0.2cm}
    \label{tab:rgbimage_results}
\end{table*}

\begin{table*}[tb]
    \centering
    \caption{Class-based pixel classification results for the \texttt{simage} mode. To generate this table all the pixels contained in the images of the testing set are considered as a single set. \rebnew{All the results are provided in percentage.}}
    \vspace{0.2cm}
    \begin{tabular}{lcccc}
        \hline
        \multicolumn{1}{c}{\bfseries Class} 
        & \multicolumn{1}{c}{\bfseries Sensitivity}
        & \multicolumn{1}{c}{\bfseries Specificity}
        & \multicolumn{1}{c}{\bfseries Accuracy}
        & \multicolumn{1}{c}{\bfseries Balanced accuracy}
        \\
        \hline
        Attached gingiva  &  40.64  &  99.48  &  97.93  &  70.06  \\
        Enamel  &  52.61  &  98.67  &  94.09  &  75.64  \\
        Hair  &  69.61  &  99.87  &  99.43  &  84.74  \\
        Hard palate  &  2.91  &  99.68  &  90.89  &  51.29  \\
        Lip  &  60.49  &  99.70  &  96.94  &  80.09  \\
        Oral mucosa  &  64.09  &  84.95  &  83.14  &  74.52  \\
        Skin  &  86.05  &  94.94  &  92.94  &  90.49  \\
        Soft palate  &  55.18  &  98.82  &  98.66  &  77.00  \\
        Tongue  &  64.58  &  98.99  &  94.36  &  81.79  \\
        \hline
        \multicolumn{1}{c}{\bfseries Average}   &  55.13  &  97.23  &  94.26  &  76.18  \\
        \end{tabular}
    \vspace{0.2cm}
    \label{tab:simage_results}
\end{table*}

\begin{table*}[tb]
    \centering
    \caption{Image-based accuracy for the different input types. The presented accuracy is the average of the images in the testing set. The accuracy for a single image is computed as the coefficient of the pixels correctly predicted divided by the total number of annotated pixels in the image. As when presenting class-based results, only the tissue classes with more than 1M pixels in the dataset have been considered. \rebnew{Accuracy results are provided in percentage.}}
    \vspace{0.2cm}
    \begin{tabular}{lc}
        \hline
        \multicolumn{1}{c}{\bfseries Input type}  &  \multicolumn{1}{c}{\bfseries Accuracy}
        \\
        \hline
        RGB pixels (\texttt{rgbpixel})           &  $39.39$  \\
        Hyperspectral pixels (\texttt{spixel})   &  $49.48$  \\
        RGB image (\texttt{rgbimage})            &  $52.51$  \\
        Hyperspectral image (\texttt{simage})    &  $54.98$  \\
\end{tabular}
    \vspace{0.2cm}
    \label{tab:image_based_results}
\end{table*}

\subsection{Evaluation protocol}
\label{sec:evaluation_protocol}
For performance testing purposes, the $215$ annotated images provided in ODSI-DB are partitioned into training ($90\%$) and testing ($10\%$)\footnote{The training/testing split is available for download at \rebmod{\url{https://synapse.org/segodsidb}}.}.
\rebnew{The training/testing split was generated randomly. We consider a training/testing split as valid when all the classes are represented in the training set (so we can perform inference on images containing any of the classes). 
In order to generate a valid training/testing split, we follow the next steps. For each class, we make a list of all the images that contain pixels of such class. We randomly pick one of those images and put it in the training set. After looping over all the classes, the training set contains at least one image with pixels of each class.
We split the remaining images into training and testing with a probability p=$0.5$.
}
\rebmod{As there are classes whose pixels can be found only in one or two images of the dataset, not all the classes are present in the testing split.
This is for example the case of the classes \textit{fibroma}, \textit{makeup}, \textit{malignant lesion}, \textit{fluorosis}, and \textit{pigmentation}.
Therefore, for reporting purposes, we ignore heavily underrepresented classes and concentrate on those tissue classes with at least $1$ million pixel samples:
\textit{skin}, \textit{oral mucosa}, \textit{enamel}, \textit{tongue}, \textit{lip}, \textit{hard palate}, 
\textit{attached gingiva}, \textit{soft palate}, and \textit{hair}.
}
In addition, we report \textit{class-based} results and \textit{image-based} results. 
The \textit{class-based} results are computed by taking all the annotated pixels contained in the testing images as a single set. A confusion matrix is then built for each class, where the positives are the pixels of such class, and the negatives are the pixels of all the other classes. Sensitivity, specificity, accuracy and balanced accuracy (arithmetic mean of sensitivity and specificity) are reported for each class. An average across classes is also reported. 
To obtain \textit{image-based} results, we compute the average accuracy across all the images of the testing set, where the accuracy of any given image is calculated as the coefficient between the number of pixels accurately classified (regardless of the class) divided by the number of pixels annotated in the image.

\subsection{Ablation study: spatial and spectral information}
\label{sec:spatial_vs_spectral}
The class-based results are shown in Tables \ref{tab:rgbpixel_results}, \ref{tab:spixel_results}, \ref{tab:rgbimage_results}, \ref{tab:simage_results} for the input modes \texttt{rgbpixel}, \texttt{spixel}, \texttt{rgbimage}, \texttt{simage}, respectively.

The class-based comparison between RGB to hyperspectral pixel inputs led to close results, except for enamel and lip classes, where hyperspectral pixels helped improve the performance by $4\,$pp and $6\,$pp, respectively. The balanced accuracy over classes showed a slight improvement of $1.1\,$pp when using multiple bands. However, when comparing the accuracy averaged over images, where better-represented classes (i.e. skin, oral mucosa, enamel, tongue, lip) have a higher weight, the hyperspectral accuracy showed an improved accuracy of $10\,$pp over RGB pixel inputs.

When comparing the class-based \texttt{rgbimage} and \texttt{simage} results, a mild improvement is observed when using the extended spectral range. The average balanced accuracy achieved was $74.39$\,\% for RGB reconstructions, and $76.18$\,\% for hyperspectral images. These results, along with the $10\,$pp gap when moving from single-pixel inputs to images suggest that without considering DL architecture changes, the spatial information is the main driver of segmentation performance in dental imaging. Nonetheless, common dental conditions such as calculus, gingiva erosion, and caries are related to two classes in particular, attached gingiva and enamel. While the enamel is distinct from the rest of the tissue, and relatively trivial to spot in an RGB image, the attached gingiva is better segmented when hyperspectral information is available, as shown by its improved balanced accuracy from $63.09$\,\% (RGB) to $70.06$\,\% (hyperspectral).

The image-based results (see Table \ref{tab:image_based_results}) display a clear performance gap ($>10\,$pp) between RGB and hyperspectral pixel inputs. This gap comes down to $2\,$pp when comparing RGB to hyperspectral images. 
\section{Conclusions}
\label{sec:conclusions}

In this work we performed an ablation study to discern how the availability of spatial and spectral information impacts the segmentation performance. We reported baseline results for four types of input data, single RGB pixels, hyperspectral pixels, RGB images and hyperspectral images. In addition, we provided an improved method to reconstruct the RGB images from the hyperspectral data provided in ODSI-DB.

We reported a mild improvement in the segmentation results on ODSI-DB when using hyperspectral information. However, the main driver of segmentation performance for the dental anatomy present in the dataset seems to be the availability of spatial information. It is when moving from pixel classification to full image segmentation that we reported the largest rise in segmentation performance.

Future work stems in several directions. An interesting research question is whether, by means of hyperspectral imaging we can mitigate the annotation effort, which is one of the current issues in the CAI field. 
That is, the mild improvement in segmentation performance achieved with hyperspectral inputs could be potentially exploited to annotate fewer images without sacrificing segmentation performance.
This would be particularly interesting for the field of hyperspectral endoscopy, as it represents an additional benefit in favour of the use of hyperspectral endoscopes. 
Another future direction is the exploration of convolutional architectures that take advantage of the hyperspectral nature of the data. Current state-of-the-art models such as U-Net have been optimised for RGB images, hence by simply replacing the input we fall on the risk of not taking full advantage of the hyperspectral information available. 
\section*{Acknowledgement(s)}
This work was supported by core and project funding from the Wellcome/EPSRC [WT203148/Z/16/Z; NS/A000049/1; WT101957; NS/A000027/1].
This study/project is funded by the NIHR [NIHR202114]. The views expressed are those of the author(s) and not necessarily those of the NIHR or the Department of Health and Social Care.
This project has received funding from the European Union's Horizon 2020 research and innovation programme under grant agreement No 101016985 (FAROS project).
TV is supported by a Medtronic / RAEng Research Chair [RCSRF1819\textbackslash7\textbackslash34].
CH is supported by an InnovateUK Secondment Scholars Grant (Project Number 75124).
For the purpose of open access, the authors have applied a CC BY public copyright licence to any Author Accepted Manuscript version arising from this submission.

\section*{Disclosure statement}
TV, ME, and SO are co-founders and shareholders of Hypervision Surgical.
TV also holds shares from Mauna Kea Technologies.

\bibliographystyle{tfcse}
\bibliography{library}

\clearpage
\appendix
\section{ODSI-DB dataset details}
\begin{table*}[htb!]
    \centering
    \caption{ODSI-DB statistics. Number of pixels per class and number of images in the dataset containing pixels of each class.}
    \vspace{0.2cm}
    \begin{tabular}{lrr}
        \hline
        \multicolumn{1}{c}{\bfseries Class} &
        \multicolumn{1}{c}{\bfseries Number of pixels} &
        \multicolumn{1}{c}{\bfseries Number of images} \\ 
        \hline
        Skin & 18993686 & 137 \\ 
        Out of focus area & 8454491 & 116 \\ 
        Oral mucosa & 7917543 & 117 \\ 
        Enamel & 4913805 & 158 \\ 
        Tongue & 4081689 & 53 \\ 
        Lip & 3920600 & 134 \\ 
        Hard palate & 2375998 & 54 \\ 
        Specular reflection & 1931845 & 179 \\ 
        Attached gingiva & 1922545 & 89 \\ 
        Soft palate & 1398594 & 43 \\ 
        Hair & 1383970 & 40 \\ 
        Marginal gingiva & 804393 & 97 \\ 
        Prosthetics & 755554 & 24 \\ 
        Shadow/Noise & 732209 & 61 \\ 
        Plastic & 255017 & 68 \\ 
        Metal & 196682 & 52 \\ 
        Gingivitis & 161874 & 39 \\ 
        Attrition/Erosion & 100919 & 36 \\ 
        Inflammation & 81098 & 10 \\ 
        Pigmentation & 43144 & 2 \\ 
        Calculus & 28615 & 27 \\ 
        Initial caries & 22008 & 30 \\ 
        Stain & 19428 & 46 \\ 
        Fluorosis & 17872 & 2 \\ 
        Microfracture & 14759 & 17 \\ 
        Root & 13962 & 17 \\ 
        Plaque & 10024 & 8 \\ 
        Dentine caries & 6616 & 16 \\ 
        Ulcer & 5552 & 11 \\ 
        Leukoplakia & 4623 & 7 \\ 
        Blood vessel & 3667 & 8 \\ 
        Mole & 2791 & 10 \\ 
        Malignant lesion & 1304 & 1 \\ 
        Fibroma & 593 & 1 \\ 
        Makeup & 406 & 2 \\ 
    \end{tabular}
    \vspace{0.2cm}
    \label{tab:class_to_numpix}
\end{table*}

\end{document}